\documentclass[12pt]{article}
\usepackage[T1]{fontenc}
\usepackage[centertags]{amsmath}
\usepackage{amsfonts}
\usepackage{amssymb}
\usepackage{amsthm}
\usepackage{newlfont}
\usepackage{epsfig}
\usepackage{amscd}

\newcommand{\bd}{{\boldsymbol d}}
\newcommand{\bu}{{\boldsymbol u}}
\newcommand{\bv}{{\boldsymbol v}}
\newcommand{\be}{{\boldsymbol e}}

\newcommand{\bA}{{\boldsymbol A}}

\newcommand{\PP}{{\mathbb P}}

\newcommand{\ZZ}{{\mathbb Z}}
\newcommand{\KK}{{\mathbb K}}
\newcommand{\FF}{{\mathbb F}}
\newcommand{\LL}{{\mathbb L}}

\newcommand{\VV}{{\mathbb V}}

\newcommand{\calC}{{\mathcal C}}

\newtheorem{Th}{Theorem}
\newtheorem{Cor}[Th]{Corollary}
\newtheorem{Lem}[Th]{Lemma}
\newtheorem{Prop}[Th]{Proposition}
\theoremstyle{definition}
\newtheorem{Def}{Definition}
\newtheorem{Ex}{Example}
\theoremstyle{remark}
\newtheorem*{Rem}{Remark}
\begin{document}
\begin{center}
{\Large The Hirota equation
 over finite fields. 
 
 Algebro-geometric approach and multisoliton solutions}
 
\bigskip

{\large A. Doliwa$^\#$, M. Bia{\l}ecki$^{\dag\ddag *}$, P. Klimczewski$^*$}

\bigskip

$^\#${\it Wydzia{\l} Matematyki i Informatyki, Uniwersytet
Warmi\'nsko--Mazurski

ul. \.Zo{\l}nierska 14A, 10-561 Olsztyn, Poland}

\bigskip

$^*${\it Instytut Fizyki Teoretycznej, Uniwersytet Warszawski

ul. Ho\.{z}a 69, 00-681 Warszawa, Poland}

\bigskip

$^\dag${\it Instytut Geofizyki PAN

ul. Ksi\c{e}cia Janusza 64, 01-452 Warszawa, Poland}

\bigskip

$^\ddag${\it Instytut Fizyki Teoretycznej, Uniwersytet w Bia{\l}ymstoku

ul. Lipowa 41, 15-424 Bia{\l}ystok, Poland} 

\end{center}

\begin{abstract}
We consider the Hirota equation (the discrete analog of the
generalized Toda system)
over a finite field. 
We present the general algebro-geometric method of construction of solutions
of the equation. As an example we construct analogs 
of the multisoliton solutions for which the wave functions and the 
$\tau$-function can be found using rational functions. Within the class of
multisoliton solutions we isolate generalized breather-type solutions
which have no direct counterparts in the complex field case.

\bigskip

\noindent PACS 2003 numbers: 02.30.Ik, 02.10.-v, 05.45.-a



\end{abstract}

\section{Introduction} 
Cellular automata are mathematical models of physical systems in
which space and time variables are discrete, and physical quantities 
take only finite number of values \cite{Wolfram}. In spite of simple
formulation they are capable to describe wide variety of phenomena, for
example traffic flow, immune systems, flow through porous media, fluid
dynamics, ferromagnetism. Due to their completely
discrete nature cellular automata
are naturally suitable for computer simulations. However in this field there
are not so many exact analytical results providing solutions with a given
global behavior. The goal of this paper is to present a general method of
construction of solutions to the cellular automaton associated with the
Hirota equation. 

The Hirota equation \cite{Hirota}, the integrable discretization of the
generalized Toda system \cite{Mikhailov}, is one of the most important 
soliton equations. 
Its various limits give rise to variety of integrable equations, moreover
it is the basic system for studying solvable quantum models
\cite{KLWZ}. 

Like many other integrable systems the Hirota equation has simple geometric
interpretation. In the paper we will use the following (geometric)
form of the Hirota equation
\begin{multline} \label{eq:Hir-geom}
\tau_m (n_1,n_2)\;\tau_m (n_1+1,n_2+1)= \\
 \tau_m(n_1+1,n_2) \; \tau_m(n_1,n_2+1) -\tau_{m-1}(n_1+1,n_2)\;
\tau_{m+1}(n_1,n_2+1). \qquad
\end{multline}
One can describe it as the equation governing the so 
called Laplace sequence of
two dimensional lattices of planar quadrilaterals \cite{DCN,Dol-Hir}. 
This interpretation can be embedded into a more general theory of 
multidimensional lattices of planar quadrilaterals and their transformations
\cite{MQL,TQL,DS-sym,Dol-IMDG}.    
It was noticed in \cite{Dol-RC,q-red} that geometric constructions in the
integrable discrete geometry (in particular, those leading to the Hirota 
equation) should
work also on the level of finite geometries \cite{Hirschfeld}, i.e., 
geometries over finite fields \cite{LidlNiederreiter}. This observation has
been developed in the present paper. 

The question of construction of integrable systems with solutions taking
values in a discrete set (soliton cellular automata) is not new and it was
undertaken in a number of papers, see for example 
\cite{BBGP,BruschiSantiniCA,TTMS,MNSTTTM}. In particular, in paper
\cite{BBGP} other Hirota equation (equivalent to the discrete sine-Gordon
equation \cite{HirotaSG}) is investigated in the context of finite fields.

In the paper we present general method of finding solutions of the
Hirota equation \eqref{eq:Hir-geom} over finite fields
by using algebro-geometric methods, standard in a complex domain
in the soliton
theory \cite{KWZ,BBEIM}. We change however the field of definition of the
underlying algebraic curves from the complex numbers to a finite field
(see also earlier
algebro-geometric papers \cite{Drinfeld,Mumford,Anderson,Thakur}
where such a possibility was considered). 

It turns out that algebraic geometry over finite fields has become recently
very important in practical use, especially in modern approaches to public
key cryptography \cite{Koblitz} and in the theory of 
error correcting codes \cite{Stichtenoth}. With respect to the last
application we would like to mention paper \cite{NM} where 
dynamics of the finite Toda molecule (a reduction of the Hirota equation) 
over finite fields was studied from the point of view of a decoding 
algorithm. 

We do not present here direct connection of the objects of the paper 
with the integrable discrete geometry over finite fields. This connection 
becomes clear
when approach to the Hirota equation presented here (see also \cite{KWZ})
is compared with results of
\cite{AKV,Dol-DTRS,Dol-IMDG}, where methods of algebraic geometry over the
field of complex numbers have been applied to construct integrable geometric
lattices. 

The layout of the paper is as follows. In Section~\ref{sec:alg-geom-Hir} we
present the general algebro-geometric scheme for construction of 
solutions of the Hirota equation. 
Section~\ref{sec:vacuum+N-soliton} is devoted to construction of
multisoliton solution on an algebraic curve starting from the vacuum
solution. Finally, in Section~\ref{sec:N-soliton} we give in explicit form 
solutions of the Hirota equation for the background algebraic curve being 
the projective line. In particular, we present the mechanism (based on the
action of the Galois group) of generation
of generalized breather-type solutions and we discuss periodicity 
of the solutions.

\section{Solutions of the Hirota equation from algebraic curves over finite
fields}
\label{sec:alg-geom-Hir}
This Section is motivated by algebro-geometric (over the complex field)
approach to the Hirota
equation (in a different form) \cite{KWZ}  and by papers 
\cite{AKV,Dol-DTRS,Dol-IMDG} on application of algebro-geometric methods
to integrable discrete geometry. It turns out that basic ideas 
of the algebro-geometric approach to soliton theory can be
transferred to the level of integrable systems in finite fields. 
The notions and results form the theory of algebraic curves over finite
fields which we use here can be found in \cite{Stichtenoth}.

Consider an algebraic projective curve, 
absolutely irreducible, nonsingular, of genus $g$, defined over the finite 
field $\KK=\FF_q$ with $q$ elements, where $q$ is a power of a
prime integer $p$. By $\calC_{\KK}$ we denote
the set of $\KK$-rational points of the curve.
By $\overline{\KK}$ denote the algebraic closure of 
$\KK$, i.e., $\overline{\KK} = \bigcup_{\ell=1}^\infty \FF_{q^\ell}$, and by
$\calC_{\overline{\KK}}$ denote the corresponding (infinite) set of
$\overline{\KK}$-rational points of the curve. 
The action of the Galois group $G(\overline{\KK}/\KK)$ (of automorphisms of 
$\overline{\KK}$ which are identity on $\KK$) extends naturally
to action on $\calC_{\overline{\KK}}$.

Let us choose:\\
1. two pairs of points $a_i,b_i\in\calC_\KK$, $i=1,2$, \\
2. $N$ points $c_\alpha\in\calC_{\overline{\KK}}$, $\alpha=1,\dots,N$,
which satisfy the following $\KK$-rationality condition
\[
\forall \sigma\in  G(\overline{\KK}/\KK), \quad 
\sigma(c_\alpha) = c_{\alpha^\prime},
\]
3. $N$ pairs of points $d_\beta, e_\beta\in\calC_{\overline{\KK}}$, 
$\beta=1,\dots,N$, which satisfy the following $\KK$-rationality condition 
\begin{equation} \label{eq:K-rat-cond}
\forall \sigma\in  G(\overline{\KK}/\KK): \quad
\sigma(\{d_\beta,e_\beta\})=\{d_{\beta^\prime},e_{\beta^\prime}\},
\end{equation}
4. $g$ points $f_\gamma\in\calC_{\overline{\KK}}$, $\gamma=1,\dots,g$,
which satisfy the following $\KK$-rationality condition
\[
\forall \sigma\in  G(\overline{\KK}/\KK), \quad 
\sigma(f_\gamma) = f_{\gamma^\prime},
\]
5. the infinity point $h_\infty\in\calC_\KK$.
\begin{Rem}
We consider here only the generic case and assume that all the points used
in the construction are generic and distinct. In particular, genericity
assumption implies that the divisor $D=\sum_{\gamma=1}^g f_\gamma$ is
non-special.
\end{Rem}
\begin{Rem}
It is enough the check the $\KK$-rationality conditions 
in any extension field $\LL\supset\KK$  
of rationality of all the points used in the construction.
\end{Rem}
\begin{Def}
Fix $\KK$-rational local parameters $t_i$ at $b_i$, $i=1,2$.
For any integers $n_1,n_2,m\in \ZZ$ define the wave function 
$\psi_{1,m}(n_1,n_2)$ as a
rational function with the following properties\\
1. it has pole of the order at most $n_1+m+1$ at $b_1$ and the pole of order
at most $n_2 - m$ at $b_2$,\\ 
2. its first nontrivial coefficient of the expansion in $t_1$ at $b_1$ is
normalized to one, \\ 
3. it has zeros of order at least $n_1$ at $a_1$, and of order at least
$n_2$ at $a_2$, \\ 
4. it has at most simple poles at points $c_\alpha$, $\alpha=1,\dots,N$, \\
5. it has zero at least
of the first order at the infinity point $h_\infty$, \\
6. it has at most simple poles at points $f_\gamma$, $\gamma=1,\dots,g$,\\
7. it satisfies $N$ constraints
\begin{equation} \label{eq:constraints}
\psi_{1,m}(n_1,n_2)(d_\beta)=\psi_{1,m}(n_1,n_2)(e_\beta), \quad
\beta=1,\dots,N.
\end{equation}
For the same set of points we define the wave function 
$\psi_{2,m}(n_1,n_2)$ as a
rational function which differs from the function $\psi_{1,m}(n_1,n_2)$ 
only in the properties $1$ and $2$ \\
1$_{2}$. it has pole of the order at most $n_1+m$ at $b_1$ and it has pole 
of the order at most $n_2-m+1$ at $b_2$,\\
2$_{2}$. its first nontrivial coefficient of the expansion in $t_2$ at $b_2$ 
is normalized to one. 
\end{Def}
\begin{Rem}
The functions $\psi_{i,m}(n_1,n_2)$ are $\KK$-rational, which follows 
from $\KK$-rationality conditions of sets of points in their definition.
\end{Rem}
\begin{Rem}
As usual, zero (pole) of a negative order means pole (zero) of the
corresponding positive order. Correspondingly one should exchange the
expressions "at most" and "at least" in front of the orders of poles and
zeros.
\end{Rem}
\begin{Prop}
The wave functions $\psi_{i,m}(n_1,n_2)$ are unique.
\end{Prop}
\begin{proof}
We show this for the first function.
By the Riemann--Roch theorem the dimension (over $\KK$) of the divisor
\[
\sum_{\alpha=1}^N c_\alpha + \sum_{\gamma=1}^g f_\gamma - h_\infty - 
n_1 a_1 - n_2 a_2 + (n_1+1+m)b_1 + (n_2 - m) b_2
\]
is equal to $N+1$. Under the
genericity assumption the $N$ constraints \eqref{eq:constraints} and the
normalization condition at $b_1$ remove the freedom.
\end{proof}
\begin{Cor}
In the next section we show, starting from the wave functions for $N=0$ one
can construct the functions for arbitrary $N$.
\end{Cor}
\begin{Rem}
To make subsequent formulas more transparent from now on we will skip 
frequently
the dependence on the parameters $n_1,n_2$.
\end{Rem}
In the generic case, which we assume in the sequel, when the order of 
the pole of $\psi_{1,m}$ at $b_1$ is
$n_1+m+1$ and the order of the pole of $\psi_{2,m}$ at $b_2$ is
$n_1-m+1$ denote by $Q_{12,m}(n_1,n_2)$ the first nontrivial
coefficient of expansion of $\psi_{1,m}$ at $b_2$, and by 
$Q_{21,m}(n_1,n_2)$ the first nontrivial
coefficient of expansion of $\psi_{2,m}$ at $b_1$, i.e.,
\[ \psi_{1,m} = \frac{1}{t_2^{n_2-m}}
\left( Q_{12,m} + \dots \right), \qquad
\psi_{2,m} = \frac{1}{t_1^{n_1+m}}
\left( Q_{21,m} + \dots \right). 
\]
\begin{Rem}
The functions $Q_{12,m}$ and $Q_{21,m}$ take values in the field $\KK$ of 
the definition of the curve.
\end{Rem}
Denote by $T_i$ the operator of translation
in the variable $n_i$, $i=1,2$,
for example $T_1 \psi_{2,m}(n_1,n_2) = \psi_{2,m}(n_1+1,n_2) $.
\begin{Prop} \label{prop:equations-psi}
The function $\psi_{1,m}$ 
satisfies equations
\begin{eqnarray}
T_2 \psi_{1,m} - \psi_{1,m} & = & 
(T_2 Q_{12,m})\psi_{2,m} , \label{eq:D2-psi1}\\  
\psi_{1,m+1} - T_1\psi_{1,m} & = &  - \frac{T_1 Q_{12,m}}{Q_{12,m}}
\psi_{1,m},  \label{eq:m+1-psi1} \\
\psi_{1,m-1} & = & \frac{1}{Q_{21,m}}
\psi_{2,m}. \label{eq:m-1-psi1}
\end{eqnarray}
The analogous system for $\psi_{2,m}$ is obtained by exchanging 
indices $1$ with $2$ and reversing the shift in the discrete variable $m$
\begin{eqnarray}
T_1 \psi_{2,m} - \psi_{2,m}& = &
(T_1 Q_{21,m})\psi_{1,m} , \label{eq:D1-psi2}\\
\psi_{2,m-1} - T_2\psi_{2,m} & = &  - 
\frac{T_2 Q_{21,m}}{Q_{21,m}} \psi_{2,m}, \label{eq:m-1-psi2}\\
\psi_{2,m+1} & = & \frac{1}{Q_{12,m}} \psi_{1,m}. \label{eq:m+1-psi2}
\end{eqnarray}
\end{Prop}
\begin{proof}
To prove the first equation \eqref{eq:D2-psi1} notice that the left
hand side has all properties of the function $\psi_{2,m}$ except of the
normalization and must be therefore proportional to $\psi_{2,m}$. The
coefficient of proportionality can be fixed comparing expansions at $b_2$.
Other equation can be proved in the same way.
\end{proof}

Fix $\KK$-rational local parameters $\tilde{t}_i$ at $a_i$, $i=1,2$. 
In the generic case when the order of $\psi_{i,m}$ at $a_i$ is $n_i$ by 
$\rho_{i,m}(n_1,n_2)$ denote first non-trivial 
coefficients of the expansion of $\psi_{i,m}$ at $a_i$, i.e.,
\[
\psi_{i,m} = \tilde{t}_i^{n_i} \left( \rho_{i,m} + \dots \right).
\]
Similarly, in the generic case when the order of $\psi_{i,m}$ at 
$a_j$, $j\ne i$, is $n_j$ by 
$\chi_{ij,m}(n_1,n_2)$ denote the first non-trivial 
coefficients of the expansion of $\psi_{i,m}$ at $a_j$, i.e.,
\[
\psi_{i,m} = \tilde{t}_j^{n_i} \left( \chi_{ij,m} + \dots \right), \quad
i\ne j.
\]
\begin{Prop} \label{prop:tau}
There exists a $\KK$-valued potential (the $\tau$-function) 
defined (up to a constant) by formulas
\begin{eqnarray}
T_1\tau_m & = & \rho_{1,m}\tau_m ,  \label{eq:rho1-tau} \\
T_2\tau_m & = & \rho_{2,m}\tau_m ,  \label{eq:rho2-tau} \\
\tau_{m+1} & = & (-1)^{n_1+n_2} Q_{12,m} \tau_m. \label{eq:Q12-tau}
\end{eqnarray}
\end{Prop}
\begin{proof}
The first terms in expansion of equations \eqref{eq:D2-psi1} and
\eqref{eq:D1-psi2} at $a_1$ give 
\begin{eqnarray}
T_2 \rho_{1,m} - \rho_{1,m} & = & (T_2Q_{12,m}) \chi_{21,m},
\nonumber \\
0 - \chi_{21,m} & = & (T_1Q_{21,m})\rho_1, \nonumber
\end{eqnarray}
which combined together give
\begin{equation} \label{eq:T2-rho1}
T_2 \rho_{1,m} = \left( 1 - (T_2Q_{12,m})(T_1Q_{21,m}) \right) \rho_{1,m}.
\end{equation}
Similarly, but changing the expansion point to $a_2$ we obtain
\[
T_1 \rho_{2,m} = \left( 1 - (T_2Q_{12,m})(T_1Q_{21,m}) \right) \rho_{2,m}.
\]
Both equations imply
\[
(T_2 \rho_{1,m}) \rho_{2,m}  = (T_1 \rho_{2,m}) \rho_{1,m}, 
\]
which is the compatibility condition of equations \eqref{eq:rho1-tau} and
\eqref{eq:rho2-tau}.

Expansion of equation \eqref{eq:m+1-psi1} at $a_1$
gives
\begin{equation} \label{eq:m+1-rho1}
\rho_{1,m+1} = - \frac{T_1Q_{12,m}}{Q_{12,m}}\rho_{1,m},
\end{equation} 
which is the compatibility condition of equations \eqref{eq:rho1-tau} and
\eqref{eq:Q12-tau}.
Finally, by comparing equations \eqref{eq:m-1-psi1} and \eqref{eq:m+1-psi2} 
we obtain
\begin{equation} \label{eq:m+1-Q21}
Q_{21,m+1} = \frac{1}{Q_{12,m}},
\end{equation}
which, combined with the following
consequence of expansion of \eqref{eq:m-1-psi2}
at $a_2$
\[
\rho_{2,m-1} = - \frac{T_2Q_{21,m}}{Q_{21,m}}\rho_{2,m},
\]
gives the compatibility condition of equations \eqref{eq:rho2-tau} and
\eqref{eq:Q12-tau}.
\end{proof}
\begin{Cor}
Equation \eqref{eq:T2-rho1} written in terms of the $\tau$-function
reads
\begin{equation}
\tau_m \;T_1T_2\tau_m = T_1\tau_m \;T_2 \tau_m -T_1\tau_{m-1}\;
T_2\tau_{m+1},
\end{equation}
which is the Hirota equation \cite{Hirota}.
\end{Cor}
\begin{Cor}
Compatibility condition of equations \eqref{eq:T2-rho1} and 
\eqref{eq:m+1-rho1}
written in terms of the function $Q_{12,m}$
reads
\begin{equation}
\frac{T_1 T_2 Q_{12,m}}{T_2 Q_{12,m}} - \frac{T_1 Q_{12,m}}{Q_{12,m}} =
\frac{T_1 T_2 Q_{12,m}}{T_1 Q_{12,m-1}} - \frac{T_2 Q_{12,m+1}}{Q_{12,m}}.
\end{equation}
\end{Cor}

\section{Construction of solutions using vacuum functions}
\label{sec:vacuum+N-soliton}
Results of this Section were motivated by papers 
on the fundamental transformation of
quadrilateral lattices in a vectorial formulation \cite{MDS,TQL,M-tau} and 
on the algebro-geometric
interpretation of this transformation \cite{Dol-DTRS,Dol-IMDG}.

In the case $N=0$ let us add superscript $0$ to all functions defined 
above and call them the vacuum functions. The functions for arbitrary $N$
can be constructed with the help of $N$ new functions, which we
define below. 
\begin{Def}
Fix local parameters $t_\alpha$ at $c_\alpha$, $\alpha=1,\dots,N$.
For any $\alpha$ define the function $\phi^0_{\alpha,m}$ by the following 
conditions:\\
1. it has pole of the order at most $n_1+m$ at $b_1$ and the pole of order
at most $n_2 - m$ at $b_2$,  \\
2 it has zeros of order at least $n_1$ at $a_1$, and of order at least $n_2$ 
at $a_2$, \\ 
3. it has at most simple pole at the point $c_\alpha$ and the first nontrivial 
coefficient of the expansion in $t_\alpha$ at $c_\alpha$ is
normalized to one, \\
4. it has zero at least of the first order at the infinity point $h_\infty$,\\
5. it has at most simple poles at points $f_\gamma$, $\gamma=1,\dots,g$.
\end{Def}
\begin{Rem}
The function $\phi^0_{\alpha,m}$ is unique but usually it is 
not $\KK$-rational. 
\end{Rem}
\begin{Lem}
Denote by $\psi^0_{i,m}(\bd,\be)$, $i=1,2$, the column with $N$ entries 
of the form
\[
\left[\psi^0_{i,m}(\bd,\be)\right]_\beta =
\psi^0_{i,m}(d_\beta)-\psi^0_{i,m}(e_\beta), \quad \beta=1,\dots,N, 
\]
denote by $\phi^0_{\bA,m}$ 
the row with $N$ entries 
\[ \left[ \phi^0_{\bA,m} \right]_\alpha = \phi^0_{\alpha,m}, \quad
\alpha=1,\dots,N, 
\]
and denote by $\phi^0_{\bA,m}(\bd,\be)$ the
$N\times N$ matrix whose element in row $\alpha$ and column $\beta$ is
\[
\left[\phi^0_{\bA,m}(\bd,\be)\right]_{\alpha \beta}=
\phi^0_{\alpha,m}(d_\beta)-\phi^0_{\alpha,m}(e_\beta) , \quad
\alpha,\beta=1,\dots,N.
\]
Then the wave functions $\psi_{i,m}$ read
\begin{equation} \label{eq:psi-i-N-soliton}
\psi_{i,m} = \psi_{i,m}^0 - \phi^0_{\bA,m} 
[\phi^0_{\bA,m}(\bd,\be)]^{-1}
\psi^0_{i,m}(\bd,\be).
\end{equation}     
\end{Lem}
\begin{proof}
Denote the right hand side of equation \eqref{eq:psi-i-N-soliton} by
$\widehat{\psi_{i,m}^0}$. By building the column 
$\widehat{\psi_{i,m}^0}(\bd,\be)$ with $N$ entries of the form 
$\widehat{\psi_{i,m}^0}(d_\beta)-\widehat{\psi_{i,m}^0}(e_\beta)$ we can
easily show that $\widehat{\psi_{i,m}^0}(\bd,\be)=0$. This demonstrates 
that the function $\widehat{\psi_{i,m}^0}$
satisfies constraints \eqref{eq:constraints}.
One can check check that $\widehat{\psi_{i,m}^0}$ satisfies also 
other properties which define uniquely the function $\psi_{i,m}$.
\end{proof}
In the generic case denote by $H^0_{i,\alpha,m}$ the first nontrivial 
coefficient of 
expansion of the function $\phi^0_{\alpha,m}$ in the 
uniformization parameter $t_i$ at $b_i$, for
example,
\[ \phi^0_{\alpha,m} = \frac{1}{t_1^{n_1+m}}\left( H^0_{1,\alpha,m} + \dots
\right).
\] 
\begin{Lem}
The corresponding expressions for $Q_{ij,m}$, $i\ne j$, and for 
$\rho_{i,m}$ read
\begin{eqnarray}
Q_{ij,m} & = & Q^0_{ij,m} -  H^0_{j,\bA,m}
[\phi^0_{\bA,m}(\bd,\be)]^{-1} \psi^0_{i,m}(\bd,\be), \quad i\ne j,
\label{eq:Q-N-soliton}\\
\rho_{i,m} & = & \rho_{i,m}^0\left( 1 + (T_iH^0_{i,\bA,m}) 
[\phi^0_{\bA,m}(\bd,\be)]^{-1} \psi^0_{i,m}(\bd,\be) \right),
\label{eq:rho-N-soliton}
\end{eqnarray}
where $H^0_{i,\bA,m}$ is
the row with $N$ entries $H^0_{i,\alpha,m}$.
\end{Lem}
\begin{proof}
Because equation \eqref{eq:Q-N-soliton} can be obtained by expansion of
formula \eqref{eq:psi-i-N-soliton} at $b_j$, only equation
\eqref{eq:rho-N-soliton} needs an explanation. 
Notice first equation
\begin{equation} \label{eq:Di-Phi}
T_i\phi_{\alpha,m}^0 - \phi_{\alpha,m}^0 = 
(T_i H^0_{i,\alpha,m})\psi^0_{i,m},
\end{equation}
which can be shown in the same way like equations of 
Proposition~\ref{prop:equations-psi}. Denote by $F^0_{i,\alpha,m}$
the first nontrivial coefficient
of expansion of $\phi^0_{\alpha,m}$ at $a_i$, for example
\[ \phi^0_{\alpha,m} = \frac{1}{\tilde{t}_1^{n_1}}
\left( F^0_{1,\alpha,m} + \dots \right).
\] 
Expansion of equation \eqref{eq:psi-i-N-soliton} at $a_i$ gives
\[
\rho_{i,m}^0(T_iH^0_{i,\alpha,m})=-F^0_{i,\alpha,m},
\]
which concludes the proof.
\end{proof}
We will use the following result, which can be proven by induction with
respect to the dimension of the vector space $\VV$.
\begin{Lem} \label{lem:det}
Given $\bu\in\VV$ and $\bv^*\in\VV^*$, if $1_\VV$ is the 
identity operator on $\VV$ then
\[ 
\det (1_\VV + \bu \otimes \bv^*) = 1 + \langle \bv^* , \bu \rangle .
\]
\end{Lem}
\begin{Prop} \label{prop:tau-N-soliton}
Using the above notation the $\tau$-function can be constructed by the
following formula
\begin{equation} \label{eq:tau-N-soliton}
\tau_{m} = \tau_{m}^0 \det \phi^0_{\bA,m}(\bd,\be).
\end{equation}     
\end{Prop}
\begin{proof}
Notice that equation \eqref{eq:Di-Phi} implies
\[
[\phi^0_{\bA,m}(\bd,\be)]^{-1} T_i \phi^0_{\bA,m}(\bd,\be) =
1_{\overline{\KK}^N} + \left( [\phi^0_{\bA,m}(\bd,\be)]^{-1} 
\psi^0_{i,m}(\bd,\be) \right) \otimes (T_iH^0_{i,\bA,m}),
\]
which gives, by Lemma \ref{lem:det},
\begin{equation} \label{eq:Ti-det}
\frac{\det T_i \phi^0_{\bA,m}(\bd,\be) }{\det \phi^0_{\bA,m}(\bd,\be)} = 
1 + (T_iH^0_{i,\bA,m})[\phi^0_{\bA,m}(\bd,\be)]^{-1} 
\psi^0_{i,m}(\bd,\be).
\end{equation}
Comparing equation \eqref{eq:Ti-det} with
equation \eqref{eq:rho-N-soliton} 
and taking into account equations \eqref{eq:rho1-tau}-\eqref{eq:rho2-tau} 
we obtain
\begin{equation} \label{eq:tau-det-i}
\frac{\det T_i \phi^0_{\bA,m}(\bd,\be) }{\det \phi^0_{\bA,m}(\bd,\be)} =
\frac{T_i \tau_m / T_i \tau^0_m}{\tau_m / \tau^0_m}.
\end{equation}

Notice the following equation
\[ 
\phi_{\alpha,m+1}^0 - \phi_{\alpha,m}^0= -
\frac{H^0_{2,\alpha,m}}{Q^0_{12,m}} \psi^0_{1,m},
\] 
which can be shown in the same way like equations of 
Proposition~\ref{prop:equations-psi}. It implies that
\[
[\phi^0_{\bA,m}(\bd,\be)]^{-1} \phi^0_{\bA,m+1}(\bd,\be) =
1_{\overline{\KK}^N} - \frac{1}{Q^0_{12,m}} 
\left( [\phi^0_{\bA,m}(\bd,\be)]^{-1} 
\psi^0_{1,m}(\bd,\be) \right) \otimes (H^0_{2,\bA,m}),
\] 
which gives 
\begin{equation} \label{eq:m-det}
\frac{\det \phi^0_{\bA,m+1}(\bd,\be) }{\det \phi^0_{\bA,m}(\bd,\be)} = 
1 - \frac{1}{Q^0_{12,m}} H^0_{2,\bA,m}[\phi^0_{\bA,m}(\bd,\be)]^{-1} 
\psi^0_{1,m}(\bd,\be).
\end{equation}
Comparing equation \eqref{eq:m-det} with
equation \eqref{eq:Q-N-soliton} 
and taking into account equation \eqref{eq:Q12-tau} we obtain
\begin{equation} \label{eq:tau-det-m}
\frac{\det \phi^0_{\bA,m+1}(\bd,\be) }{\det \phi^0_{\bA,m}(\bd,\be)} =
\frac{ \tau_{m+1} / \tau^0_{m+1}}{\tau_m / \tau^0_m},
\end{equation}
which, together with equation \eqref{eq:tau-det-i}, concludes the proof.
\end{proof}
\begin{Cor}
Notice that equation \eqref{eq:tau-N-soliton} is valid up to a
(nonessential) change of 
the initial value of the $\tau$-function, which is due to introduction 
of the integration constant from formulas \eqref{eq:tau-det-i} and 
\eqref{eq:tau-det-m}.
\end{Cor}
\begin{Cor}
Starting with $\KK$-valued function $\tau^0_m$ and the local
parameters $t_\alpha$ at $c_\alpha$ chosen in a consistent way with the
action of the Galois group $G(\overline{\KK}/\KK)$ 
on $\calC_{\overline{\KK}}$ we obtain $\KK$-valued function $\tau_m$. 
\end{Cor}
\section{Multisoliton solutions}
\label{sec:N-soliton}
We present here explicit formulas for the vacuum functions in the simplest
case $g=0$. Then with the help of these expressions we present some examples
of $N$-soliton solutions. In constructing the vacuum functions
we will use the standard parameter $t$ on the 
projective line $\PP(\KK)$ and we put $h_\infty=\infty$.

Explicit forms of the wave functions read
\begin{eqnarray}
\psi^0_{1,m} &= &\frac{1}
{(t-b_1)^{n_1+1+m}}
\frac{(t-a_1)^{n_1} (t-a_2)^{n_2}(b_1-b_2)^{n_2-m}}
{(b_1-a_1)^{n_1}(b_1-a_2)^{n_2}(t-b_2)^{n_2-m}}, \nonumber \\
\psi^0_{2,m} &=& \frac{1}
{(t-b_2)^{n_2+1-m}}
\frac{(b_2-b_1)^{n_1+m}(t-a_1)^{n_1} (t-a_2)^{n_2}}
{(t-b_1)^{n_1+m}(b_2-a_1)^{n_1}(b_2-a_2)^{n_2}}, \nonumber
\end{eqnarray}    
which gives formulas for the functions $Q^0_{12,m}$ and $Q^0_{21,m}$
\begin{eqnarray}
Q^0_{12,m} &= & \frac{(-1)^{n_2-m}}{(b_2-b_1)^{n_1-n_2+1+2m}}
\frac{(b_2-a_1)^{n_1} (b_2-a_2)^{n_2}}
{(b_1-a_1)^{n_1}(b_1-a_2)^{n_2}}, \nonumber \\
Q^0_{21,m} &=& \frac{(-1)^{n_1+m}}{(b_1-b_2)^{n_2-n_1+1-2m}}
\frac{(b_1-a_1)^{n_1} (b_1-a_2)^{n_2}}
{(b_2-a_1)^{n_1}(b_2-a_2)^{n_2}}, \nonumber
\end{eqnarray}    
and for the functions $\rho_{1,m}$ and $\rho_{2,m}$
\begin{eqnarray}
\rho^0_{1,m} &= & \frac{(-1)^{n_1}}{(a_1-b_1)^{2n_1+1+m}}
\frac{(a_1-a_2)^{n_2}(b_1-b_2)^{n_2-m}}
{(b_1-a_2)^{n_2}(a_1-b_2)^{n_2-m}}, \nonumber \\
\rho^0_{2,m} &=& \frac{(-1)^{n_2}}{(a_2-b_2)^{2n_2+1-m}}
\frac{(b_2-b_1)^{n_1+m}(a_2-a_1)^{n_1}}
{(a_2-b_1)^{n_1+m}(b_2-a_1)^{n_1}}. \nonumber
\end{eqnarray}    
Explicit form of the vacuum $\tau$-function reads
\[ 
\tau^0_m = \frac{(-1)^{[n_1(n_1-1) + n_2(n_2-1) + m(m+1)]/2}}
{(a_1-b_1)^{n_1(n_1+m)} (a_2-b_2)^{n_2(n_2-m)}}
\frac{(a_1-a_2)^{n_1n_2}(b_1-b_2)^{(n_2-m)(m+n_1)}}
{(b_1-a_2)^{n_2(n_1+m)}(a_1-b_2)^{n_1(n_2-m)}}.
\] 
The functions $\phi^0_{\alpha,m}$, $\alpha=1,\dots,N$
have the form
\[
\phi_{\alpha,m}^0= \frac{1}{t-c_\alpha}
\frac{(t-a_1)^{n_1} (t-a_2)^{n_2}(c_\alpha-b_1)^{n_1+m}
(c_\alpha-b_2)^{n_2-m}}
{(c_\alpha-a_1)^{n_1} (c_\alpha-a_2)^{n_2}(t-b_1)^{n_1+m}(t-b_2)^{n_2-m}},
\]
and can be used, due to Proposition~\ref{prop:tau-N-soliton}, to construct the
$\tau$-function for arbitrary $N$.
 
Let $\LL=\FF_{q^\ell} \subset\overline{\KK}$ be a
field of rationality
of all the points used in the construction. Recall \cite{Lang-alg}
that if $q=p^k$ then
the Galois group
$G(\LL/\KK)$ is the cyclic group of order $\ell$ and is 
generated by $\sigma_F^k$, where $\sigma_F$ is the Frobenius automorphism
of $\LL$ defined as $\sigma_F(a)=a^p$. Therefore possible 
systems of 
points $c_\alpha$
and pairs $\{d_\beta,e_\beta\}$ can be grouped into $\KK$-rational
clusters (orbits of the group $G(\LL/\KK)$)
of the lengths being divisors of $\ell$. In the standard
nomenclature the clusters of length one correspond to solitons, and clusters
of length two give rise to breathers (the position of poles $c_\alpha$
of the wave functions must be symmetric with respect to the complex
conjugation). In finite fields we encounter new types of solutions 
(without direct analogs in the complex field case) which come
from clusters of lengths greater then two. The analog of the breather
solution will be presented in Example~\ref{ex:1}. Let us call the
$N$-soliton solution of order $\ell$ the $\KK$-rational $N$-soliton
solution with parameters in extension of $\KK$ of order $\ell$. In this
terminology the standard $N$-soliton solutions are of order one, while
the breather solution is a $2$-soliton solution of order two. 
A $3$-soliton solution of order three and a $2$-soliton solution of order 
four
are presented in Examples~\ref{ex:2} and~\ref{ex:3}. We remark that the
above terminology is not completely distinctive. 

Notice that the variables $n_1$, $n_2$ and $m$ enter exponentially
in the functions $\tau^0_m$ and $\phi^0_{\alpha,m}$ 
This implies that the $\tau$-function
is periodic in $n_1$, $n_2$ 
and $m$ with the periods being divisors of $q^\ell - 1$, which is the order
of the cyclic multiplicative group $\LL_*$.

Finally, we present the examples. For any example we describe the
fields $\KK$ and $\LL$ giving first the numbers 
$q=p^k$ and $\ell$ and then writing down the polynomial $w(x)$ 
over $\FF_p$ used to construct multiplication in the field $\LL$. 
We represent elements of $\LL$ as elements of the vector space $\FF_p^{kl}$. 
Then we give the points used in the construction of the solution 
of the Hirota equation presenting also
the action of the Galois group $G(\LL/\KK)$ on them. 

\begin{Ex} \label{ex:1} A breather solution of the Hirota equation in
$\FF_5$. Parameters of the solution take values in extension $\FF_{5^2}$ of
$\FF_5$ by the polynomial $w(x)=x^2+x+1$. The corresponding Galois group 
reads $G(\FF_{5^2}/\FF_5)=\{ id, \sigma\}$, where $\sigma^2=id$. The
parameters of the solution are chosen as follows 

$a_1=(00)$, $a_2=(02)$, $b_1=(01)$, $b_2=(04)$

$c_1=(10)$, $c_2=\sigma(c_1)=(44)$,

$d_1=(11)$, $d_2=\sigma(d_1)=(40)$,
 
$e_1=(13)$, $e_2=\sigma(e_1)=(42)$.

This solution is presented in Figures~\ref{fig:2b-m1} 
and~\ref{fig:2b-m2and5}. The elements of
$\FF_5$ are represented by:
\leavevmode\epsfysize=0.35cm\epsffile{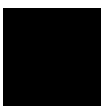} -- $(00)$, 
\leavevmode\epsfysize=0.35cm\epsffile{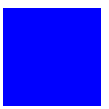} -- $(01)$, 
\leavevmode\epsfysize=0.35cm\epsffile{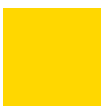} -- $(02)$,
\leavevmode\epsfysize=0.35cm\epsffile{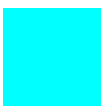} -- $(03)$, 
\leavevmode\epsfysize=0.35cm\epsffile{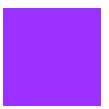} -- $(04)$.

The periods in variables $n_1$, $n_2$ and $m$ are
$12$, $24$ and $24$, correspondingly. Notice that figure for $m=5$ can be
obtained from the figure for $m=1$ by a shift in $n_2$ by four.  
\begin{figure}
\begin{center}
\leavevmode\epsfysize=9cm\epsffile{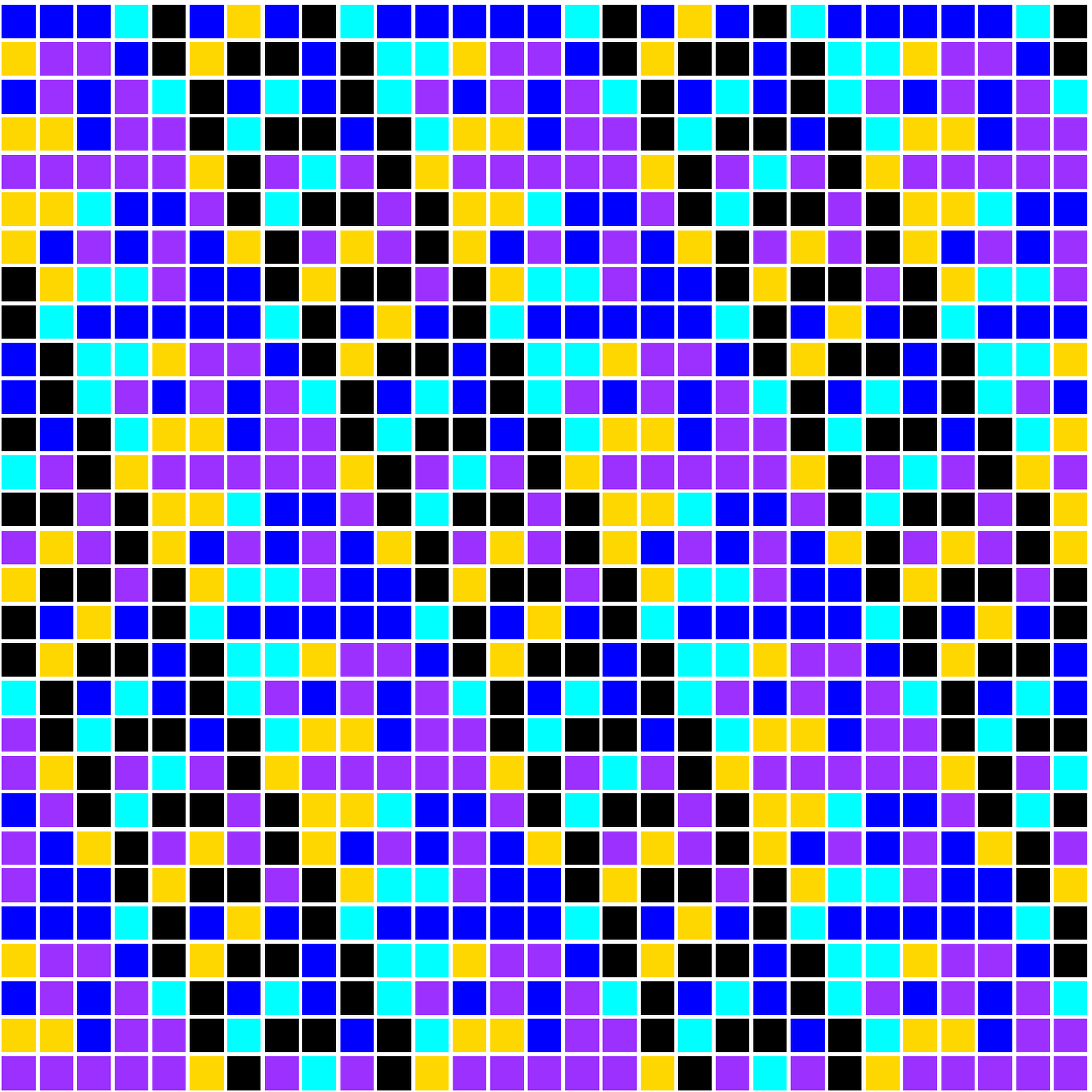}
\end{center}
\caption{
A breather solution of the Hirota equation in $\FF_5$ described
in Example~\ref{ex:1}; $n_1$ range from $0$ to $28$ (directed to the 
right), $n_2$ range from $0$ to $28$ (directed up), $m=1$.
}
\label{fig:2b-m1}
\end{figure}
\begin{figure}
\begin{center}
\leavevmode\epsfysize=7.5cm\epsffile{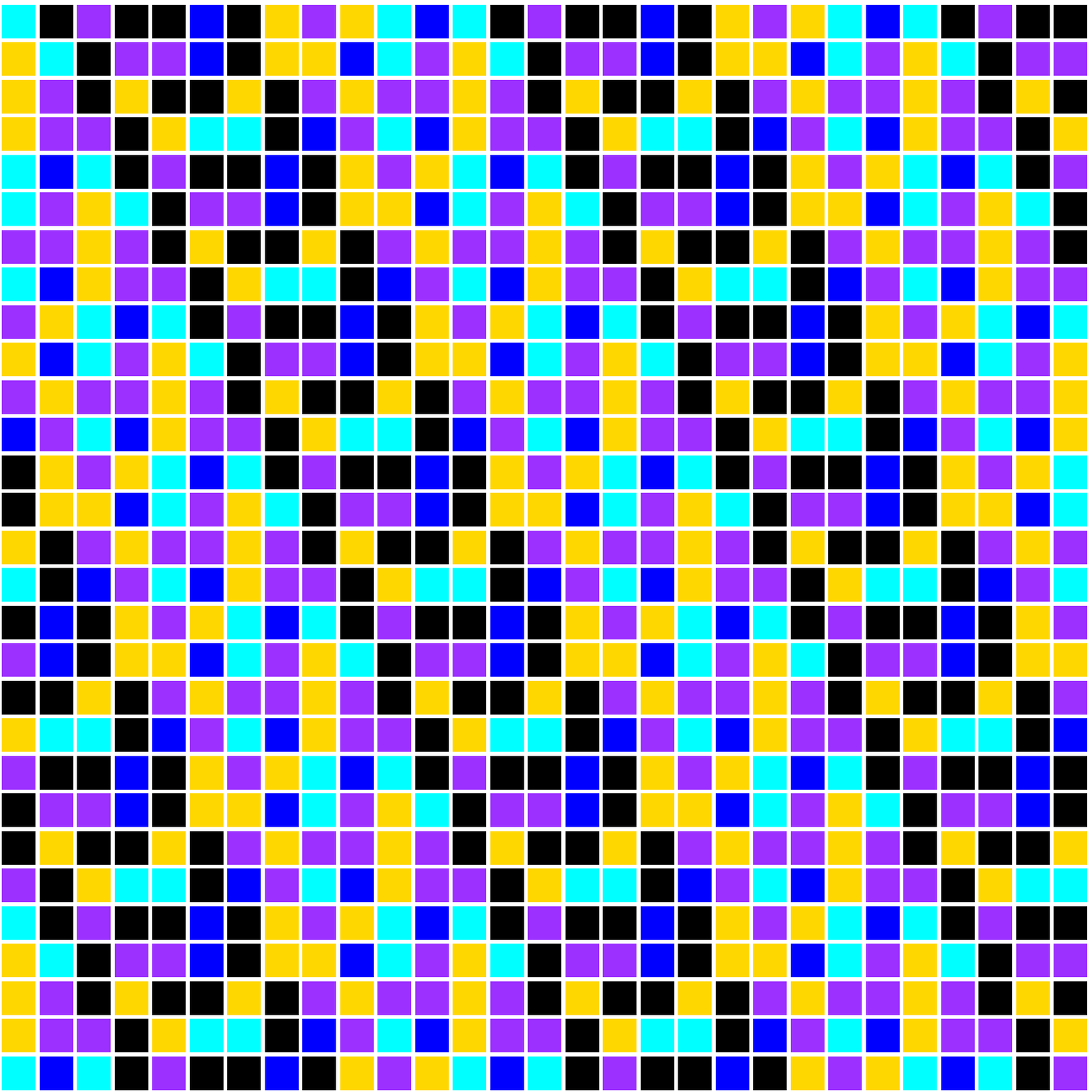}
\leavevmode\epsfysize=7.5cm\epsffile{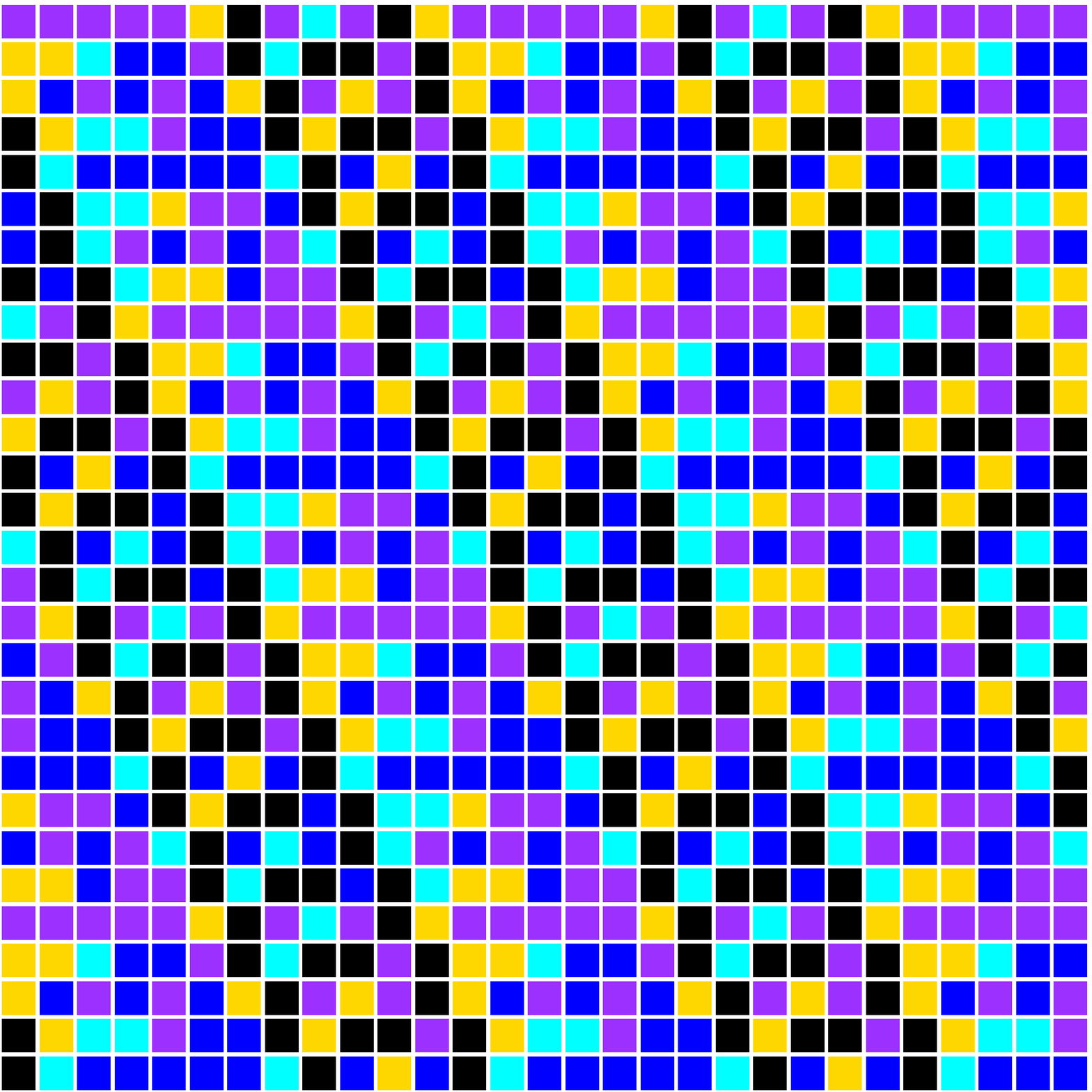}
\end{center}
\caption{A breather solution as in Figure \ref{fig:2b-m1} for 
$m=2$ and $m=5$.}
\label{fig:2b-m2and5}
\end{figure}
\end{Ex}  
\begin{Ex} \label{ex:2} A $3$-soliton solution of order $3$ of the Hirota 
equation in
$\FF_5$. This solution is the first one in the generalized breather class, 
a new type not
present in the standard complex field case. Parameters of the 
solution take values in extension $\FF_{5^3}$ of
$\FF_5$ by the polynomial $w(x)=x^3+x^2+1$. The corresponding Galois group 
reads $G(\FF_{5^3}/\FF_5)=\{ id, \sigma, \sigma^2 \}$, where $\sigma^3=id$. 
The parameters of the solution are chosen as follows: 

\nopagebreak 
$a_1=(004)$, $a_2=(003)$, $b_1=(002)$, $b_2=(001)$

\nopagebreak 
$c_1=(022)$, $c_2=\sigma(c_1)=(120)$, $c_3=\sigma^2(c_1)=(412)$,

\nopagebreak 
$d_1=(020)$, $d_2=\sigma(d_1)=(123)$, $d_3=\sigma^2(d_1)=(410)$,

\nopagebreak 
$e_1=(010)$, $e_2=\sigma(e_1)=(314)$, $e_3=\sigma^2(e_1)=(230)$.

This solution is presented in Figures~\ref{fig:3b-m0-m4-m62}
and~\ref{fig:3b-m0-130}. The elements of
$\FF_5$ are represented like in Example~\ref{ex:1}. 
From the figures one can deduce that the periods in all variables 
$n_1$, $n_2$ and $m$ are maximal and equal to $124$. 
\begin{figure}
\begin{center}
\leavevmode\epsfysize=4.5cm\epsffile{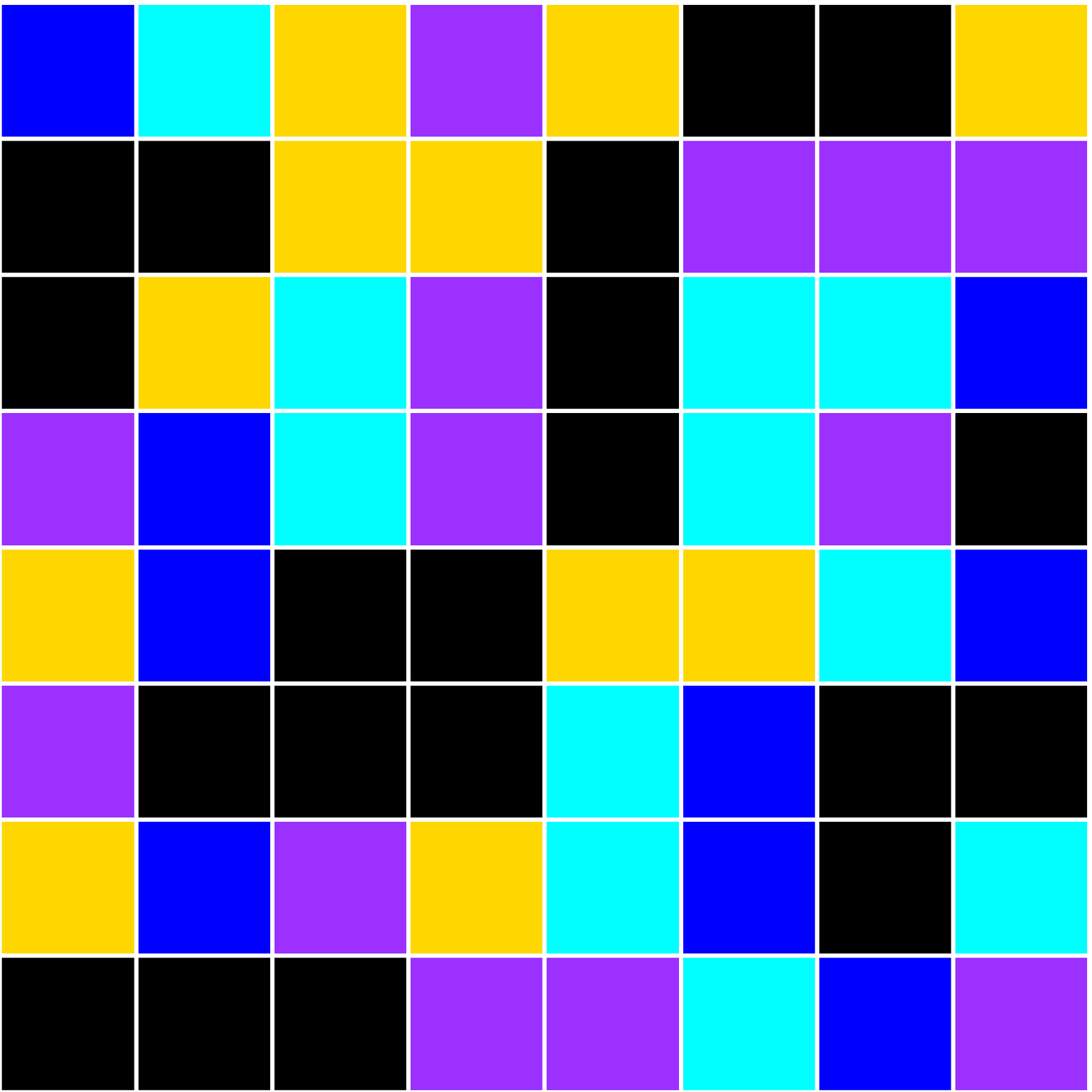}
\leavevmode\epsfysize=4.5cm\epsffile{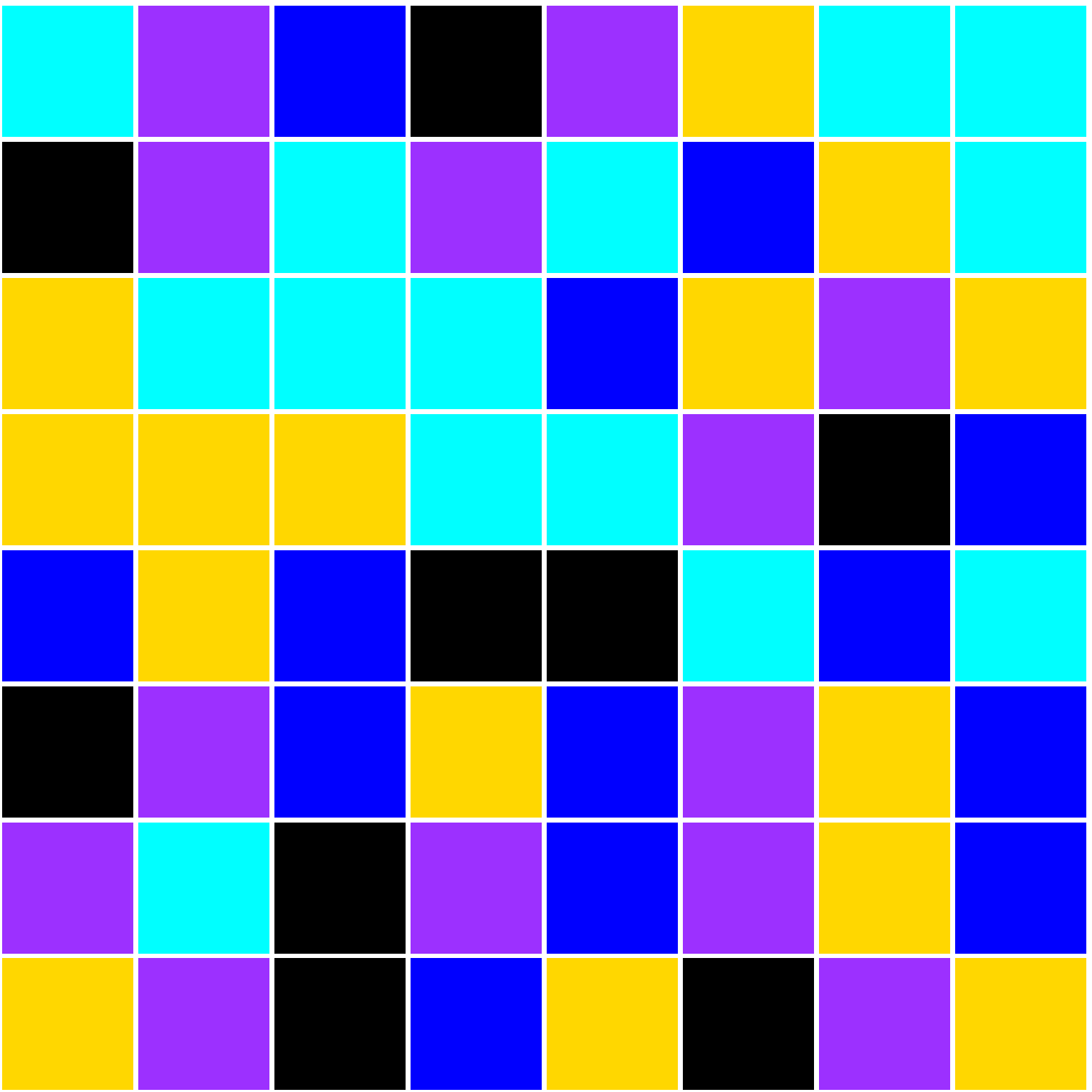}
\leavevmode\epsfysize=4.5cm\epsffile{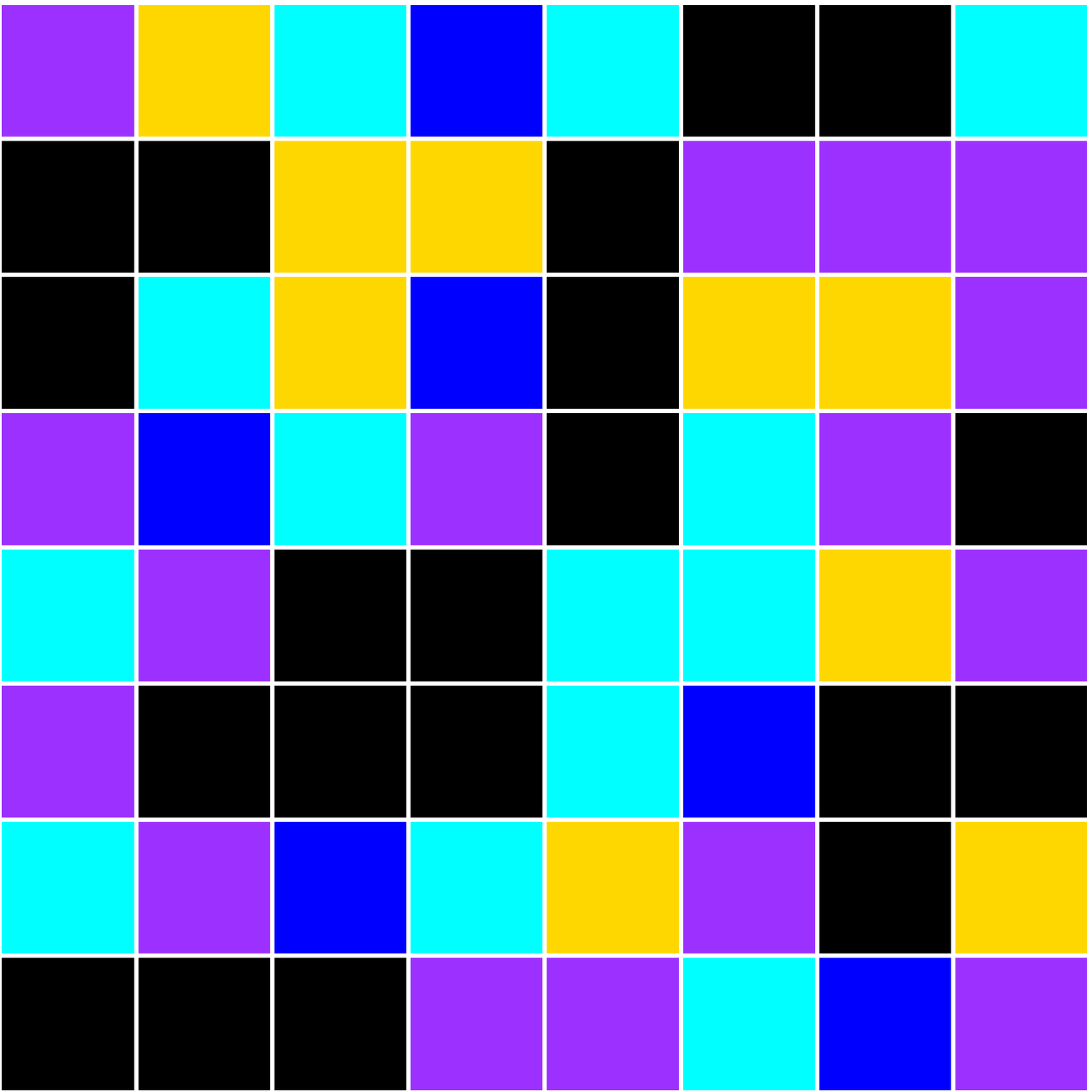}
\end{center}
\caption{A breather of order $3$
solution of the Hirota equation in $\FF_5$ described
in Example~\ref{ex:2}; $n_1$ range from $0$ to $8$ (directed to the 
right), $n_2$ range from $0$ to $8$ (directed up), $m=0$,~$4$~and $62$.}
\label{fig:3b-m0-m4-m62}
\end{figure} 
\begin{figure}
\begin{center}
\leavevmode\epsfysize=15cm\epsffile{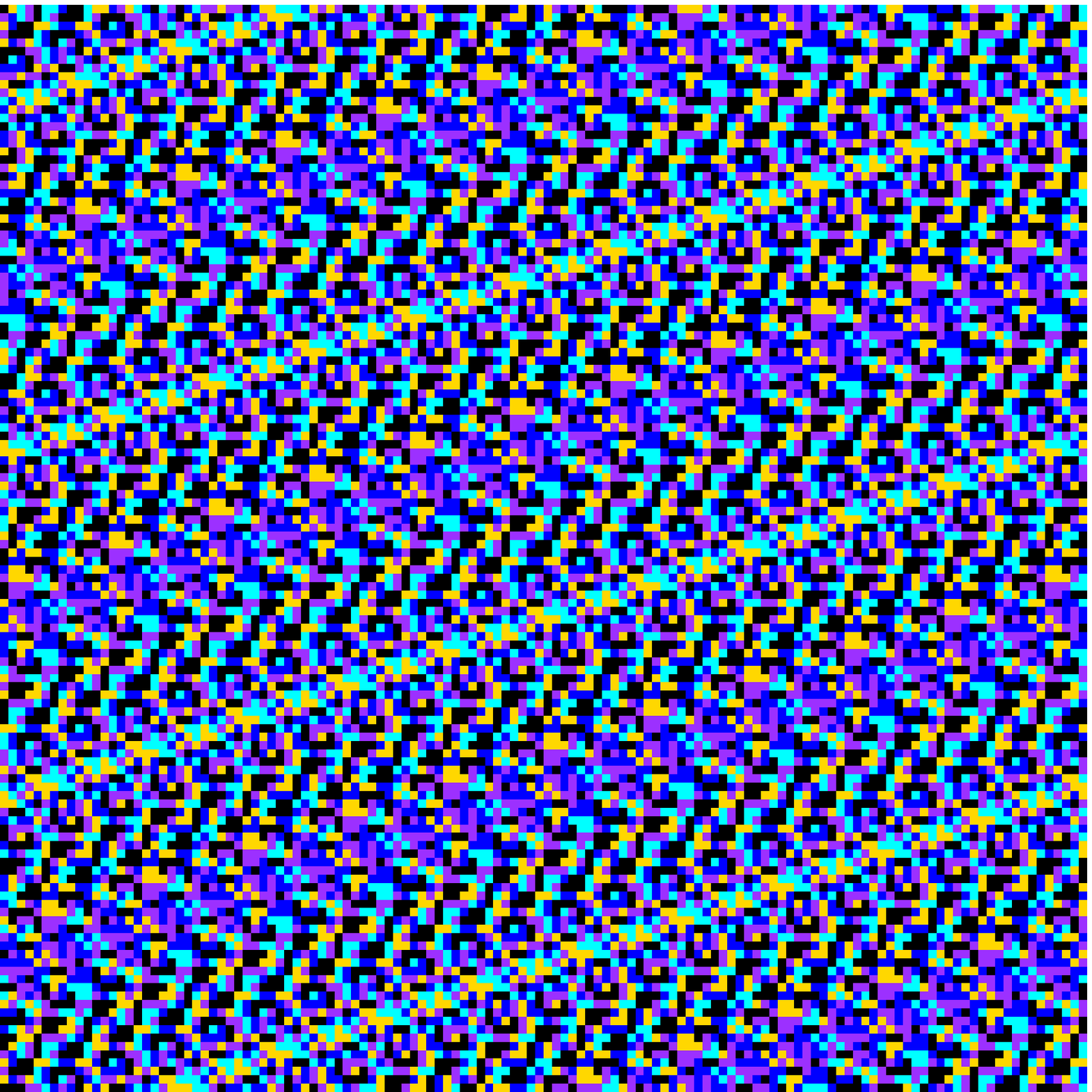}
\end{center}
\caption{
A "global" view of the first picture ($m=0$) of the  generalized
breather solution
presented in Figure \ref{fig:3b-m0-m4-m62}; $n_1$ and $n_2$ range from 
$0$ to $129$.
}
\label{fig:3b-m0-130}
\end{figure}
\end{Ex}  
\begin{Ex} \label{ex:3} The goal of this example is to present a solution in
a "small" field $\FF_4$ obtained from parameters taking values in a
relatively bigger field $\FF_{256}$. The field $\FF_{256}$ is chosen as
extension of $\FF_2$ by the polynomial $w(x)=x^8+x^6+x^5+x+1$.
The elements $(00000000)$, $(00000001)$,
$(11110000)$ and $(11110001)$ of $\FF_{256}=\FF_{2^8}$ form a subfield
isomorphic to $\FF_4$. The Galois group is generated by 
$\sigma = \sigma_F^2$, and reads
$G(\FF_{256}/\FF_4)=\{ id, \sigma, \sigma^2, \sigma^3 \}$, where 
$\sigma^4=id$. The parameters of the solution are as follows:

$a_1=(00000000)$, $a_2=(00000001)$, $b_1=(11110000)$, $b_2=(11110001)$

$c_1=(00010010)$, $c_2=\sigma(c_1)=(11100011)$,

$d_1=(00001010)$, $d_2=\sigma(d_1)=(00001001)$,
 
$e_1=\sigma^2(d_1)=(00011000)$, $e_2=\sigma^3(d_1)=(11101010)$.

Here the points $c_1$ and $c_2$ are chosen
from a subfield of $\FF_{256}$ isomorphic to $\FF_{16}$ and form a 
cluster of length two. The points $c_1$,
$c_2$, $d_1$ and $d_2$ form a cluster of length four in a way compatible
with the $\FF_4$-rationality condition \eqref{eq:K-rat-cond}. We obtain
therefore a $2$-soliton solution of order four, which also has no direct
counterpart in the complex field case. 

The solution is presented in Figure~\ref{rys:2sF256-m0-60} for $m=0$; 
The elements of $\FF_4$ are represented by:
\leavevmode\epsfysize=0.35cm\epsffile{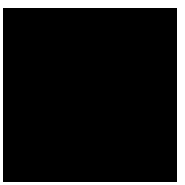} -- $(00000000)$, 
\leavevmode\epsfysize=0.35cm\epsffile{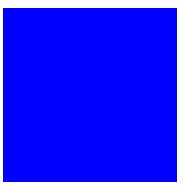} -- $(00000001)$, 
\leavevmode\epsfysize=0.35cm\epsffile{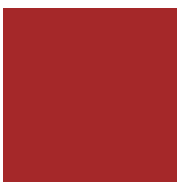} -- $(11110000)$,
\leavevmode\epsfysize=0.35cm\epsffile{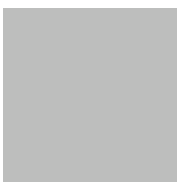} -- $(11110001)$.
The period in all variables is the same and equals $51$.
\begin{figure}
\begin{center}
\leavevmode\epsfysize=15cm\epsffile{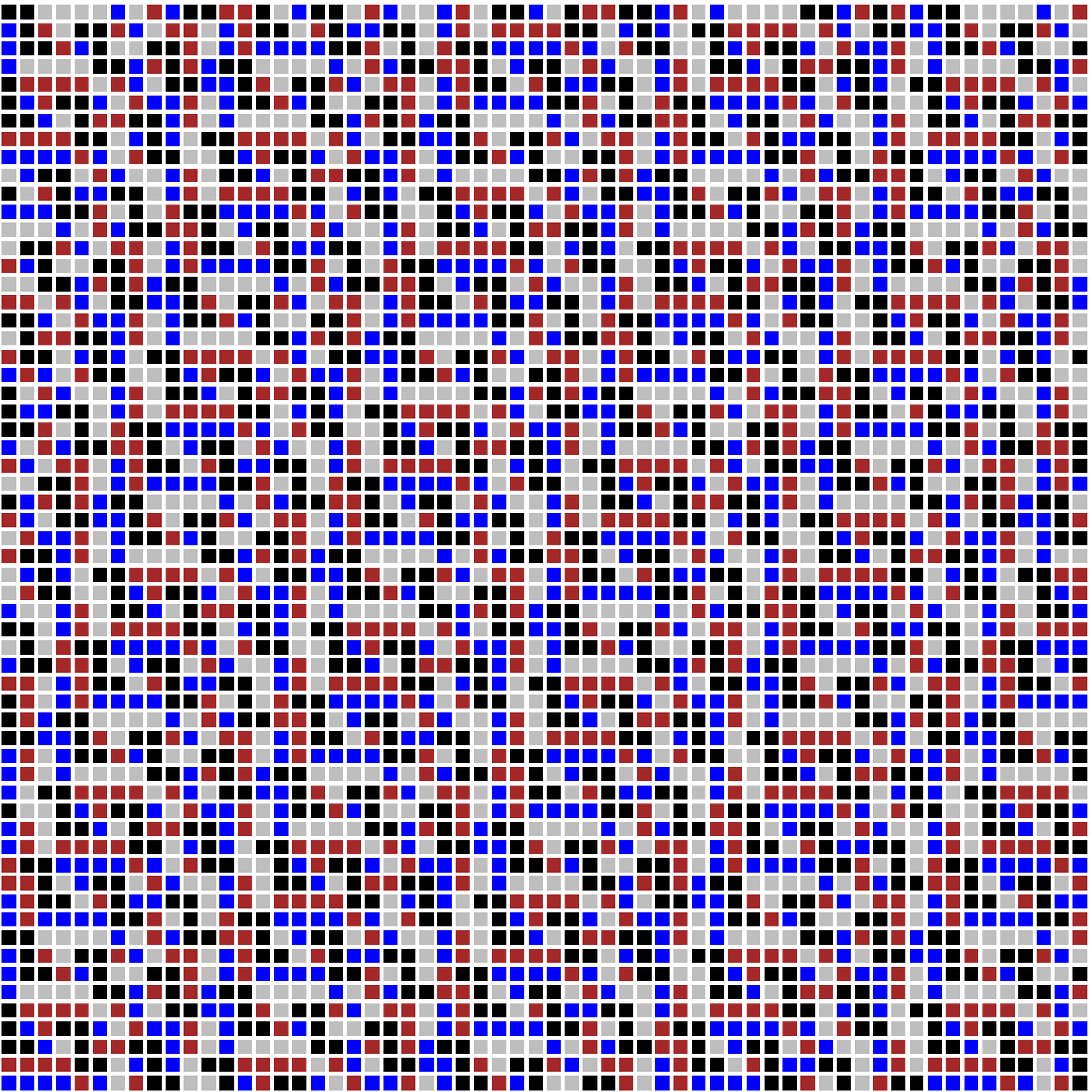}
\end{center}
\caption{
A $2$-soliton solution of order four of the Hirota equation
in $\FF_4$; $n_1$ and $n_2$ range from 
$0$ to $59$, $m=0$.
}
\label{rys:2sF256-m0-60}
\end{figure}
\end{Ex}

\section{Conclusion and remarks}
In the paper, motivated by recent developments of integrable discrete
geometry, we presented algebro-geometric method of construction of
solutions of the Hirota equation over finite fields. It turns out that the
main ideas used for integrable systems over the field of complex
numbers, e.g., application of the Riemann-Roch theorem, can be 
transferred to the level of finite fields without essential modifications.
We would like to stress that although finite fields consist of finite 
number of elements the
corresponding algebraic curves have infinite number of points (taking into
account the algebraic completion of the field) which gives rise to infinite
families of solutions of the equation. 

We have presented examples of pure 
(for the simplest algebraic curve being the
projective line) multisoliton solutions of the Hirota equation. Less
trivial examples which use techniques on Jacobians of algebraic curves will
be the subject of another paper \cite{BialDol-hyp}.

\section*{Acknowledgments}
The paper was 
partially supported by KBN grant no. 2P03B12622.

\bibliographystyle{amsplain}

\begin{thebibliography}{10}

\bibitem{AKV}
A.~A. Akhmetshin, I.~M. Krichever, and Y.~S. Volvovski, \emph{Discrete analogs
  of the {D}arboux--{E}goroff metrics}, {\tt hep-th/9905168}.

\bibitem{Anderson}
G.~W. Anderson, \emph{Rank one elliptic {$A$}-modules and {$A$}-harmonic
  series}, Duke Math. J. \textbf{73} (1994), 491--542.

\bibitem{BBEIM}
E.~D. Belokolos, A.~I. Bobenko, V.~Z. Enol'skii, A.~R. Its, and V.~B. Matveev,
  \emph{Algebro-geometric approach to nonlinear integrable equations},
  Springer-Verlag, Berlin, 1994.

\bibitem{BialDol-hyp}
M.~Bia{\l}ecki and A.~Doliwa, in preparation.

\bibitem{BBGP}
A.~Bobenko, M.~Bordemann, Ch. Gunn, and U.~Pinkall, \emph{On two integrable
  cellular automata}, Comm. Math. Phys. \textbf{158} (1993), 127--134.

\bibitem{BruschiSantiniCA}
M.~Bruschi and P.~M. Santini, \emph{Cellular automata in 1+1, 2+1 and 3+1
  dimensions, constants of motion and coherent structures}, Physica D
  \textbf{70} (1994), 185--209.

\bibitem{DCN}
A.~Doliwa, \emph{Geometric discretisation of the {Toda} system}, Phys. Lett. A
  \textbf{234} (1997), 187--192.

\bibitem{Dol-RC}
\bysame, \emph{Discrete integrable geometry with ruler and compass}, Symmetries
  and Integrability of Difference Equations (P.~Clarkson and F.~Nijhoff, eds.),
  Cambridge University Press, 1999, pp.~122--136.

\bibitem{q-red}
\bysame, \emph{Quadratic reductions of quadrilateral lattices}, J. Geom. Phys.
  \textbf{30} (1999), 169--186.

\bibitem{Dol-Hir}
\bysame, \emph{Lattice geometry of the {H}irota equation}, SIDE III --
  Symmetries and Integrability of Difference Equations (D.~Levi and
  O.~Ragnisco, eds.), CMR Proceedings and Lecture Notes, vol.~25, AMS,
  Providence, 2000, pp.~93--100.

\bibitem{Dol-DTRS}
\bysame, \emph{The {D}arboux-type transformations of integrable lattices}, Rep.
  Math. Phys. \textbf{48} (2001), 59--66.

\bibitem{Dol-IMDG}
\bysame, \emph{Integrable multidimensional discrete geometry: Quadrilateral
  lattices, their transformations and reductions}, Integrable Hierarchies and
  Modern Physical Theories (H.~Aratyn and A.~S. Sorin, eds.), Kluwer,
  Dordrecht, 2001, pp.~355--389.

\bibitem{MQL}
A.~Doliwa and P.~M. Santini, \emph{Multidimensional quadrilateral lattices are
  integrable}, Phys. Lett. A \textbf{233} (1997), 365--372.

\bibitem{DS-sym}
\bysame, \emph{The symmetric, {D}-invariant and {E}gorov reductions of the
  quadrilateral lattice}, J. Geom. Phys. \textbf{36} (2000), 60--102.

\bibitem{TQL}
A.~Doliwa, P.~M. Santini, and M.~Ma{\~n}as, \emph{Transformations of
  quadrilateral lattices}, J. Math. Phys. \textbf{41} (2000), 944--990.

\bibitem{Drinfeld}
V.~Drinfeld, \emph{Commutative subrings of some noncommutative rings},
  Functional Anal. Appl \textbf{11} (1977), 9--12.

\bibitem{HirotaSG}
R.~Hirota, \emph{Nonlinear partial difference equations. {III}. {D}iscrete
  sine-{G}ordon equation}, J. Phys. Soc. Jpn. \textbf{43} (1977), 2079--2086.

\bibitem{Hirota}
\bysame, \emph{Discrete analogue of a generalized {Toda} equation}, J. Phys.
  Soc. Jpn. \textbf{50} (1981), 3785--3791.

\bibitem{Hirschfeld}
J.~W.~P. Hirschfeld, \emph{Projective geometries over finite fields}, Clarendon
  Press, Oxford, 1998.

\bibitem{Koblitz}
N.~Koblitz, \emph{Algebraic aspects of cryptography}, Springer, Berlin, 1998.

\bibitem{KLWZ}
I.~M. Krichever, O.~Lipan, P.~Wiegmann, and A.~Zabrodin, \emph{Quantum
  integrable models and discrete classical {H}irota equations}, Commun. Math.
  Phys. \textbf{188} (1997), 267--304.

\bibitem{KWZ}
I.~M. Krichever, P.~Wiegmann, and A.~Zabrodin, \emph{Elliptic solutions to
  difference non-linear equations and related many body problems}, Commun.
  Math. Phys. \textbf{193} (1998), 373--396.

\bibitem{Lang-alg}
S.~Lang, \emph{Algebra}, Addison-Wesley, Reading, Mass., 1970.

\bibitem{LidlNiederreiter}
R.~Lidl and H.~Niederreiter, \emph{Introduction to finite fields and their
  applications}, Univ. Press, Cambridge, 1986.

\bibitem{M-tau}
M.~Ma{\~n}as, \emph{Fundamental transformations for quadrilateral lattices:
  first potentials and $\tau$-functions, symmetric and pseudo-{E}gorov
  reductions}, J. Phys A: Math. Gen. \textbf{34} (2001), 10413--10421.

\bibitem{MDS}
M.~Ma{\~n}as, A.~Doliwa, and P.~M. Santini, \emph{Darboux transformations for
  multidimensional quadrilateral lattices. {I}}, Phys. Lett. A \textbf{232}
  (1997), 99--105.

\bibitem{Mikhailov}
A.~V. Mikhailov, \emph{Integrability of a two-dimensional generalization of the
  {T}oda chain}, JETP Lett. \textbf{30} (1979), 414--418.

\bibitem{MNSTTTM}
S.~Moriwaki, A.~Nagai, J.~Satsuma, T.~Tokihiro, M.~Torii, D.~Takahashi, and
  J.~Matsukidaira, \emph{$2+1$ dimensional soliton cellular automaton},
  Symmetries and Integrability of Difference Equations (P.~Clarkson and
  F.~Nijhoff, eds.), Cambridge University Press, 1999, pp.~334--342.

\bibitem{Mumford}
D.~Mumford, \emph{An algebro-geometric construction of commuting operators and
  of solutions to the {T}oda lattice equation, {K}orteweg--de {V}ries equation,
  and related nonlinear equations}, Proceedings of the {I}nternational
  {S}ymposium on {A}lgebraic {G}eometry (M.~Nagata, ed.), Kinokuniya, Tokyo,
  1978, pp.~115--153.

\bibitem{NM}
Y.~Nakamura and A.~Mukaihira, \emph{Dynamics of the finite {T}oda molecule over
  finite fields and a decoding algorithm}, Phys. Lett. A \textbf{249} (1998),
  295--302.

\bibitem{Stichtenoth}
H.~Stichtenoth, \emph{Algebraic function fields and codes}, Springer-Verlag,
  Berlin, 1993.

\bibitem{Thakur}
D.~S. Thakur, \emph{Integrable systems and number theory in finite
  characterictic}, Physica D \textbf{152-153} (2001), 1--8.

\bibitem{TTMS}
T.~Tokihiro, D.~Takahashi, J.~Matsukidaira, and J.~Satsuma, \emph{From soliton
  equations to integrable cellular automata through a limiting procedure},
  Phys. Rev. Lett. \textbf{76} (1996), 3247--3250.

\bibitem{Wolfram}
S.~Wolfram, \emph{Theory and application of cellular automata}, World
  Scientific, Singapore, 1986.

\end{thebibliography}
\providecommand{\bysame}{\leavevmode\hbox to3em{\hrulefill}\thinspace}

\end{document}